# Near-Field Vibrational Energy Transfer for Mid-Infrared Upconversion in Plasmonic Nanogaps

Avisekh Pal[1‡], Anju Sajan[1], Christopher Sumner[1], Eman Alharbi[1§], Wolfgang Theis[1] and Rohit Chikkaraddy*[1]

[1] School of Physics and Astronomy, University of Birmingham, Birmingham, UK

**Abstract**

**Förster energy transfer underpins modern photonics, yet establishing an analogous vibrational pathway in the mid-infrared (MIR) remains highly challenging, as sub-picosecond intramolecular vibrational redistribution (IVR) suppresses intermolecular coupling. Here we demonstrate vibrational donor-acceptor transfer in the MIR and subsequent upconversion to visible luminescence enabled by sub-2 nm plasmonic nanogaps. The extreme lateral field confinement in metal-molecule-metal ring cavities defined by self-assembled molecular spacers couple efficiently to in-plane molecular dipoles. Continuous-wave MIR excitation selectively populates -C≡N vibrational donors, and plasmon-enhanced near-field coupling transfers this energy to nearby electronic acceptors, generating anti-Stokes visible emission under low power densities. Upconversion efficiencies exceeding 0.3% are observed, limited by competition between the plasmon-mediated transfer rate and IVR. These results show that extreme plasmonic confinement can redirect molecular vibrational relaxation pathways, opening a route toward vibrational nanophotonics, intermolecular interactions for bioimaging and room-temperature MIR detection based on molecular degrees of freedom.**

---

Current Address:

‡ Centre for Nanoscience and Engineering, Indian Institute of Science, Bengaluru 560012, India

§ Department of Physics, College of Science, Qassim University, Buraydah 52571, Saudi Arabia

Förster and Dexter mechanisms establish an energy transfer pathway between quantum emitters in electronically excited states, where a donor molecule non-radiatively populates electronic states of an acceptor via near-field dipole-dipole coupling[1–4]. This paradigm has enabled light harvesting, Förster resonance energy transfer (FRET) biosensing, and excitonic transport in molecular solids[5–9]. The essential requirement is that the donor lifetime exceeds or competes with the transfer rate, allowing coherent or incoherent energy migration before internal decay.

Although molecular vibrations possess well-defined dipolar transitions in the 3–30μm window, their excited-state lifetimes are typically limited to picoseconds by ultrafast intramolecular vibrational redistribution (IVR) and phonon-mediated relaxation[10–13]. This rapid relaxation suppresses intermolecular vibrational energy transfer under ambient conditions, preventing the formation of a vibrational analogue of FRET. As a result, vibrational excitations are traditionally regarded as local probes rather than mobile energy carriers. Mid-infrared (MIR) spectroscopy therefore remains fundamentally absorptive: energy is dissipated into heat before it can be transferred or converted. The development of vibrational donor-acceptor transfer has important consequences for MIR photonics and bioimaging[14–16]. Unlike the visible regime, where excitonic energy migration and photon upconversion can be engineered using molecular architectures, MIR photons cannot be efficiently converted into higher-energy emission without invoking nonlinear optical crystals or multiphoton processes. Conventional upconversion approaches include parametric frequency mixing, difference-frequency generation, entangled photon conversion, or multiphoton absorption demands high peak intensities and phase matching in bulk nonlinear media[17–21]. These constraints limit scalability, continuous-wave operation, and integration with nanoscale sensing platforms.

Recent advances in plasmonics have demonstrated that extreme electromagnetic confinement can dramatically modify spontaneous emission, vibrational absorption, and light-matter coupling strengths. Sub-10 nm metal–insulator–metal (MIM) nanogaps achieve mode volumes approaching $10^{-6}\lambda^3$ and Purcell factors exceeding $10^5$-$10^{10}$, enabling single-molecule Raman scattering, emission from dark-states, single-photon emission and vibration-mediated upconversion[22–26]. Atomic-layer-defined coaxial apertures and nanogaps have been shown to support epsilon-near-zero modes and slow-light Fabry–Pérot resonances with uniform field enhancement across sub-nanometre volumes[27–33]. These developments suggest that plasmonic cavities may provide a route to overcome the intrinsic lifetime limitation of vibrational excitations by accelerating and redirecting their relaxation pathways.

Here we demonstrate a vibrational donor-acceptor mechanism in the MIR enabled by molecularly defined plasmonic nanogaps. In our system (Fig.1a,b), a resonantly excited molecular vibration functions as a MIR energy donor, while an electronically emissive dye molecule serves as an acceptor. The MIR photon excites a vibrational quantum $|0\rangle_D \rightarrow |1\rangle_D$ in the donor molecule. This excitation is transferred to the acceptor molecule, resulting in ground vibrational population in $|1\rangle_A$. In the presence of a near-infrared (NIR) pump, the vibrational population is remotely pumped into electronic absorption in the acceptor molecule ($|1\rangle_A \rightarrow |0'\rangle_A$). This transition now relaxes to the ground state ($|0'\rangle_A \rightarrow |0\rangle_A$), exhibiting photoluminescence at wavelengths shorter than the NIR pump, resulting in anti-Stokes photoluminescence (aS-PL) and forming a mid-infrared vibrationally assisted luminescence (MIRVAL) signal. The energy of NIR pump is below the electronic absorption band ($|0\rangle_A \rightarrow |0'\rangle_A$) of the acceptor molecule, so direct transition from the ground state is energetically forbidden. In free space, the vibrational donor would relax non-radiatively via IVR on picosecond timescales, requiring ultrafast vibrational spectroscopy to probe[34,35]. However, when confined inside a deeply sub-nanometre plasmonic cavity, the vibrational excitation is coupled to a localised gap plasmon mode whose electromagnetic density of states exceeds free-space values by several orders of magnitude. This cavity-enhanced near-field mediates ultrafast dipole-dipole energy transfer from the vibrational donor to the electronic acceptor before vibrational relaxation occurs, resulting in continuous-wave anti-Stokes visible emission.

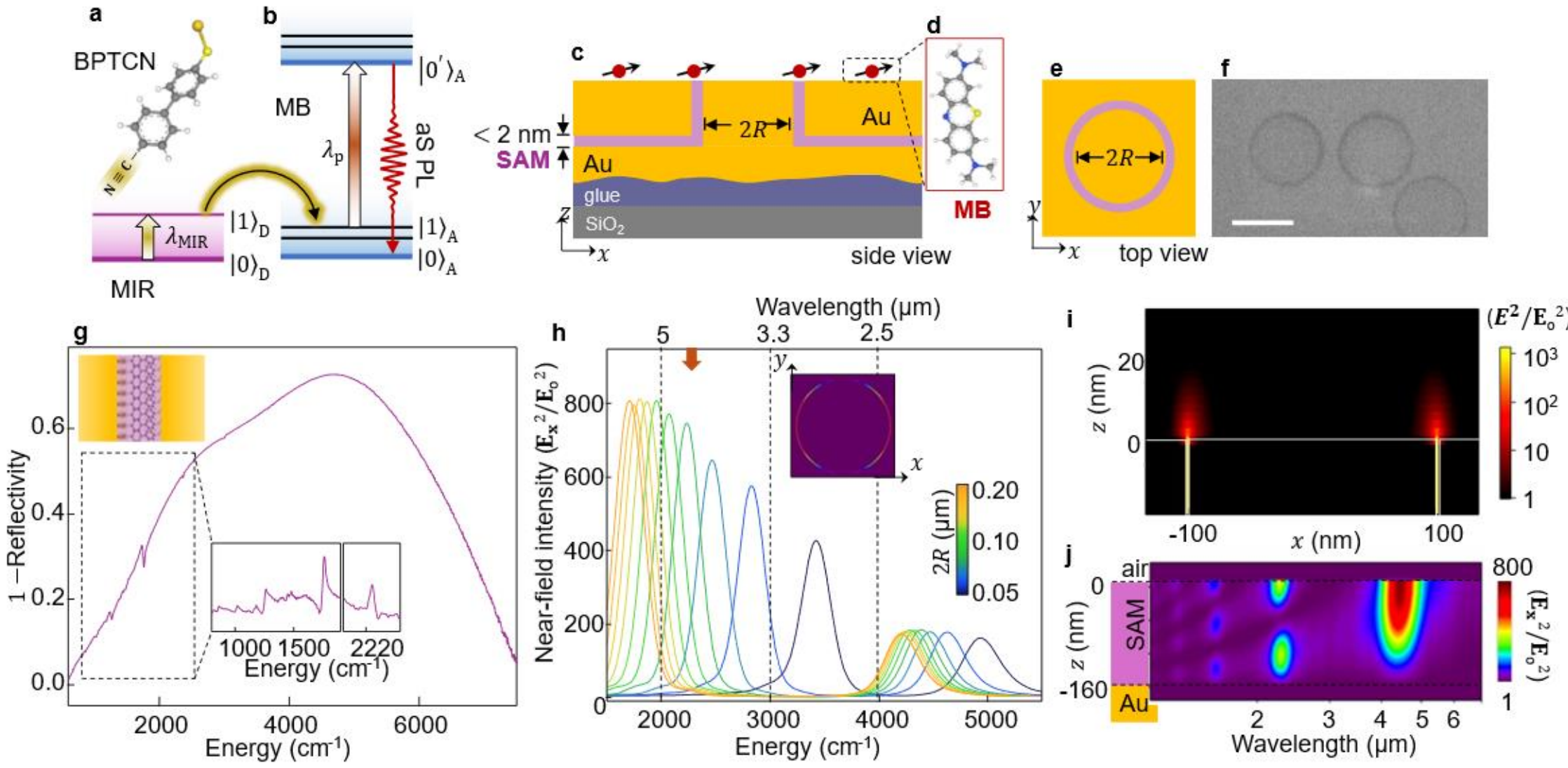

**Figure 1: Donor-acceptor vibrational energy transfer enabling MIR upconversion.** (**a**) Energy-level diagram illustrating MIR excitation ($\lambda_{\mathrm{MIR}}$ = 4.55μm) of a BPTCN self-assembled monolayer (SAM), driving the -C≡N stretching vibration from the ground state $|0\rangle_{\mathrm{D}}$ to the first vibrationally excited state $|1\rangle_D$. (**b**) The vibrational energy transfer from MIR-excited BPTCN to a nearby methylene blue (MB) molecule, leading to electronic excitation ($|1\rangle_A \rightarrow |0'\rangle_{\mathrm{A}}$) and upconverted emission ($|0'\rangle_A \rightarrow |0\rangle_{\mathrm{A}}$). (**c**) Cross-sectional schematic of the metal–insulator–metal (MIM) ring structure comprising Au film separated by a <2nm SAM spacer, with methylene blue (MB) molecules located at the metal surface. (**d**) Chemical structure of MB. (**e**) Top-view schematic of a MIM ring cavity highlighting the annular gap. (**f**) Scanning electron microscopy image of fabricated MIM rings with a diameter of 25.4±0.8μm (scale bar: 25 μm). (**g**) Fourier-transform infrared (FTIR) spectrum of MIM rings ($2R$=500nm) assembled with a monolayer of BPTCN. Inset: background-removed vibrational absorption peak of BPTCN. (**h**) Full-wave 3D simulations of the wavelength-dependent electric near-field intensity enhancement in MIM nanorings with varying ring radius $R$. The downward arrow marks $\lambda_{\mathrm{MIR}}$, resonant with the -C≡N stretching vibration of the donor (BPTCN). Inset: top-view field distribution at resonance, showing confined $E_x^2$ gap field intensity along the annular cavity. (**i**) Cross-sectional electric-field intensity map illustrating exponential field decay above the film and (**j**) vertically confined plasmonic gap modes within the SAM spacer region as a function of wavelength highlighting the cavity modes ($m_{\theta\phi}$=(10)-(13)).

**Results**

**Molecularly defined plasmonic nanogaps as vibrational energy-transfer platforms**

To construct such sub-10nm plasmonic nanocavities with resonance tuned to the MIR and to precisely orient acceptor and donor molecules along the optical fields, we used bottom-up nanoassembly. The plasmonic geometry we developed creates of metal-insulator-metal nanoring cavities using a lithography-free microsphere-assisted template-stripping process (Fig.1c-f, Supplementary Fig.S1 and S2). Briefly, monodisperse polystyrene microspheres (2R=500nm and 30µm) are first self-assembled on a silicon substrate to define the lateral dimensions of the cavities (Supplementary Fig.S3, S4). A gold film is thermally evaporated over the microspheres, coating both the substrate and the upper hemisphere of each sphere. The microspheres are subsequently removed by toluene dissolution followed by oxygen plasma etching, leaving circular apertures in the gold film with diameters defined by the microsphere size. The patterned gold surface is then functionalised with a self-assembled monolayer of 4-mercaptobiphenyl carbonitrile (BPTCN). The thiol anchor chemisorbs to gold, forming a dense monolayer spacer layer whose thickness (<2 nm) precisely defines the eventual nanogap. A second gold layer is deposited by thermal evaporation, completing the metal-molecule-metal stack. Epoxy-assisted template stripping transfers the bilayer structure onto a substrate, exposing ultrasmooth gold surfaces and embedding coaxial ring cavities with sub-2 nm gaps defined solely by the molecular monolayer thickness. The resulting fabrication yield is ∼80% over 1 $mm^2$ (Supplementary Fig.S5), determined by the fraction of microspheres that are fully removed during the lift-off, and plasma-etching steps.

The completed MIM rings support mid-infrared (MIR) gap-plasmon resonances engineered to spectrally overlap the -C≡N vibrational stretching mode (≈2220 $cm^{-1}$) of the BPTCN donor molecule. Infrared spectroscopy reveals the broad plasmon resonance of the MIM rings for 2R=500nm (Fig.1g), consistent with the designed spectral overlap. To verify molecular integrity and visible field confinement, we performed SERS mapping on BPTCN assembled within the nanogap (Supplementary Fig. S6). The observed vibrational signatures of -C≡N, both in the Stokes and anti-Stokes spectra, confirm that the BPTCN molecules remain intact throughout the fabrication process, despite being assembled prior to the second Au deposition (Supplementary Fig.S7). We note that during the second Au deposition, the high kinetic energy of incoming Au atoms could potentially damage molecules located outside the protected vertical nanogap regions and form a vertical Au bridges between the two metal layers.

Following fabrication, the acceptor molecules of methylene blue (MB) emitters were introduced by solution adsorption, yielding sub-monolayer coverage on the Au/SAM surface. MB adsorbs on Au primarily via $\pi$-metal interactions and electrostatic forces, without strong external alignment constraints. However, the lone pair on the central nitrogen can induce a slight tilt in the molecular orientation. Despite this, the lowest-energy electronic transition dipole moment lies predominantly parallel to the metal surface ($\boldsymbol{\mu}_{\mathbf{x}}$). This orientation is crucial because the gap modes supports strong in-plane electric fields ($\boldsymbol{E_x}$) localised within the molecular spacer. Full-wave simulations with plane wave-excitation reveal that the fundamental gap-plasmon mode of the nanoring is governed by azimuthal propagation around the circumference combined with vertical confinement across the sub-2 nm spacer (Fig.1h-j). The electromagnetic field distribution is characterised by $m_{\theta\phi}$ mode number with number of nodes along radial ($\theta$) and azimuthal ($\phi$) angles. The fundamental mode ($m = (10)$) maximizes confinement and local density of optical states (LDOS), yielding near-field intensity enhancements $|E/E_x|^2 \sim 8 \times 10^2$ for a $2R$=200 nm ring at $\lambda \approx$4.5µm. The resonance red-shifts with increasing $R$ as the effective index approaches saturation, while higher-order modes ($m \geq (11)$) exhibit increased radiative damping and reduced field confinement. Importantly, although the vertical MIM geometry enforces sub-nanometre capacitive confinement, the dominant electric field component inside the gap is lateral ($E_x$), parallel to the metal interfaces and directed across the slot (Supplementary Fig.S8). The light–matter interaction strength scales as $g \propto \boldsymbol{\mu} \cdot \mathbf{E}$, where $\boldsymbol{\mu}$ is the molecular transition dipole (vibrational and electronic). For both the -C≡N bond vibration in BPTCN and the electronic transition, in MB molecules oriented predominantly in-plane, $\boldsymbol{\mu}_\mathbf{x} \parallel E_x$, maximise excitation efficiency and Purcell enhancement.

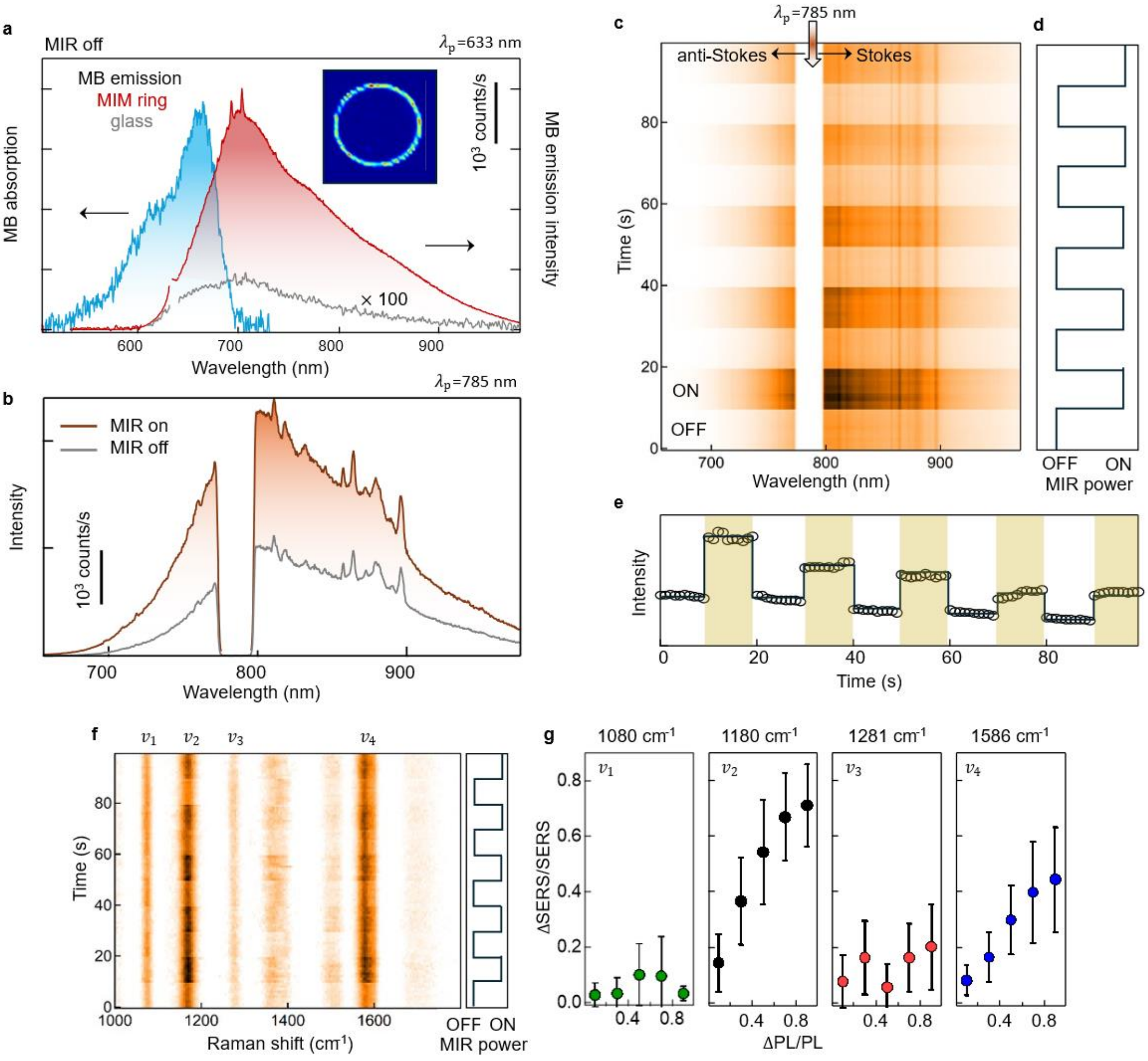

**Figure 2: Donor-acceptor mediated MIR-to-visible upconversion in MIM rings.** (a) Photoluminescence (PL) spectrum of MB from an individual MIM ring under direct optical excitation at 633nm (25µW/µm$^2$; red trace). Compared to the absorption spectrum of MB (blue) and the PL

spectrum of MB assembled on a glass substrate (grey, ×100 for clarity). The inset displays the spatial emission map, confirming that the MB emission is localized to the annular nanocavity. (**b**) Stokes and anti-Stokes emission under near-infrared excitation (785nm; 132μW/μm$^2$) with and without MIR illumination ($P_{\mathrm{MIR}}$=2mW/μm$^2$). (**c**) Temporal evolution of Stokes and anti-Stokes PL over time where the MIR illumination is periodically switched on and off (**d**). The Stokes spectrum also reveals the vibrational lines of BPTCN molecules. (**e**) Extracted anti-Stokes PL intensity as a function of time. Light-brown shaded regions correspond to MIR ON periods. (**f**) PL-subtracted SERS spectra (from Fig.2c) revealing distinct vibrational modes of the BPTCN donor molecule ($v_1$-$v_4$), collected over time as the MIR power is modulated (right panel). (**g**) Correlation between normalised SERS intensity modulation of donor vibrational modes (BPTCN: $v_1$-$v_4$) to PL modulation of the electronic acceptor (MB) and error bars indicate the standard deviation.

The extreme nanogaps in MIM rings enhance the light emission of MB molecules. We characterised the photoluminescence (PL) of MB on the plasmonic nanorings, in the absence of MIR illumination (Fig.2a). A 633nm CW laser was used to pump the samples (the energy of 633nm is above MB direct absorption); we observed a strong emission from MB on a MIM ring with $2R$=500nm, centered around ~680nm with a broad bandwidth (>100nm). The emission intensity was >300× greater than the MB monolayer on a flat glass slide (Fig.2a). This enhanced emission is attributed to the combined effects of field localisation and the Purcell effect in the nanogap cavities[36,37]. To confirm the enhancement of PL from the nanogaps, we performed confocal mapping on MIM rings ($2R$=30μm, Supplementary Fig.S9). The PL signal in the nanogaps is >150× larger than the area inside and outside the MIM rings. The effective enhancement of the PL in the nanogaps is more than >$10^{10}$ (>$10^8$ normalised to glass) after accounting for the number of molecules and the area of the electromagnetic hot spot. Indicating the localisation of visible light in extreme nanogaps where participation of high-order plasmonic modes significantly alters the emitter linewidths[38,39] and boosts emission intensity.

**Near-field vibrational energy transfer and assisted upconversion**

To perform the MIR-induced anti-Stokes upconversion, the MIM nanoring was illuminated simultaneously by the MIR laser and the NIR laser (Supplementary Fig.S10). A NIR laser at 785 nm was used to pump the samples (the energy of 785 nm is below the MB's absorption gap). The MIR laser tuned to 4.55 μm (2222cm$^{-1}$) to resonantly excite the -C≡N vibrations of BPTCN. We emphasise that all measurements were performed with CW excitation sources at low power densities (MIR: 0.5μW/μm$^2$, NIR: 100μW/μm$^2$ and MIM rings with $2R$=500nm), thus with conditions far more modest than typical nonlinear upconversion schemes. MB does not directly absorb either the MIR or NIR laser as these are not in resonance with vibrational or electronic excitations.

Under simultaneous NIR (785nm) and MIR excitation, a pronounced anti-Stokes photoluminescence (aS-PL) signal emerges (Fig. 2b), which is absent under NIR illumination alone. The MIR beam is resonant with the -C≡N stretching mode of the BPTCN spacer, enabling selective vibrational excitation within the nanogap. Subtraction of the NIR-only spectrum from the dual-excitation spectrum reveals a broadband increase in emission intensity extending on both anti-Stokes and Stokes sides of the NIR excitation line (Supplementary Fig.S11). This MIR vibrationally assisted luminescence (MIRVAL) has strongest enhancement (MIR on/ MIR off from Fig. 2b) of >550% is observed near the 680 nm matching the MB emission peak and reflects vibrationally mediated population transfer into the electronically excited manifold of MB. The MIR-induced upconversion is fully reversible and follows on/off modulation of the MIR beam (Fig.2c-e), with less than 7% decrease in total aS-PL intensity over the timeseries measurements, indicating some photobleaching. The broadband nature of the enhancement is consistent with coupling to low-$Q$ higher-order plasmonic modes that modify the local density of optical states across a wide spectral window. In addition, nanoscale roughness at the ring edges produces localised hotspots for NIR excitation that reshape the excitation and emission spectra, as we observe with direct 633nm excitation.

To resolve the microscopic pathway of energy transfer, we correlate the modulation depth of individual BPTCN vibrational modes with concurrent PL fluctuations (Fig. 2f). The PL subtracted Stokes intensities of four Raman-active modes ($v_1$-$v_4$) were extracted and compared to the underlying MB emission variation at each spectral position (Fig. 2g). The C-H in-plane bending mode at 1180cm$^{-1}$ ($v_2$) and the ring stretching mode[40] at 1586cm$^{-1}$ ($v_4$) exhibit strong correlation with MIR-induced PL modulation, whereas the C-H rocking mode at 1080cm$^{-1}$ ($v_1$) and the coupled ring stretch at 1281cm$^{-1}$ ($v_3$) show significantly weaker correlation. These mode-selective correlations indicate that energy initially absorbed into the resonant -C≡N stretch redistributes through specific intramolecular pathways before coupling to MB. Such vibrational mode-dependent modulated response is incompatible with uniform equilibrium heating, suggesting that non-equilibrium, mode-specific energy flow dominates the observed behaviour.

We quantify the MIRVAL efficiency ($\%\zeta_{\mathrm{MIRVAL}}$) by comparing the MIR-induced aS-PL intensity ($I_{aS\text{-}PL_{\lambda_{785}+\lambda_{\mathrm{MIR}}}} - I_{aS\text{-}PL_{\lambda_{785}}}$) to the photoluminescence generated under direct 633nm excitation ($I_{PL_{\lambda_{633}}}$) without MIR:

$$\%\zeta_{\mathrm{MIRVAL}} = \frac{I_{aS\text{-}PL_{\lambda_{785}+\lambda_{\mathrm{MIR}}}} - I_{aS\text{-}PL_{\lambda_{785}}}}{I_{PL_{\lambda_{633}}}} \times 100$$

All spectra were acquired at the same nanoring location and normalised for laser power and integration time. We obtain $\%\zeta_{\mathrm{MIRVAL}}$= 0.33±0.14% for BPTCN gaps. This value represents the fraction of vibrationally injected population that is converted into visible emission relative to direct electronic excitation and exceeds the expected Boltzmann population of the vibrational level ($2.6\times10^{-5}$) by ×100 at 300K and therefore cannot arise from equilibrium heating alone. While the precise microscopic role of collective -C≡N oscillators remain to be fully resolved, the observed mode-selective correlations and reversible modulation indicate vibrationally mediated near-field energy transfer as the operative mechanism.

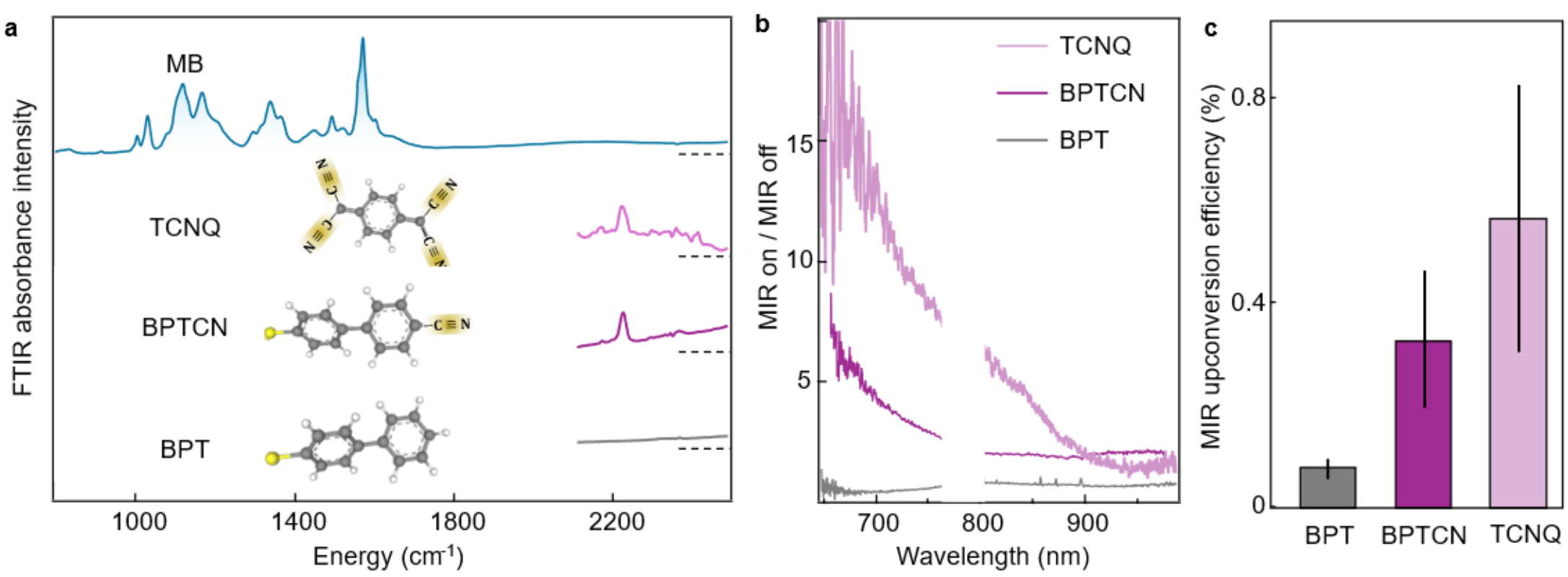

**Figure 3: Role of -C≡N in vibrational energy transfer.** (**a**) FTIR absorbance spectra of MB and three candidate donor molecules: BPTCN, BPT and TCNQ, drop-cast on a gold-coated silicon substrate. Both BPTCN and TCNQ exhibit a -C≡N stretching mode in the silent spectral zone (2000-2500cm$^{-1}$), whereas BPT lacks this vibrational feature. (**b**) MIR-induced upconversion spectra for MIM rings functionalized with different donor SAMs: TCNQ (pink), BPTCN (purple), and BPT (grey). (**c**) Histogram of extracted MIR upconversion efficiencies for the three donor molecules and error bars indicate the standard error.

To isolate the role of -C≡N vibrational donor in mediating energy transfer, we fabricated MIM rings with chemically similar biphenyl-4-thiol (BPT) as a SAM spacer instead of BPTCN. BPT forms comparable Au-thiol monolayers but does not contain a -C≡N stretch[41], and is used as a non-resonant spacer with similar gap size as BPTCN. In addition, we prepared MIM rings incorporating tetracyanoquinodimethane (TCNQ), which contains four -C≡N groups and adsorbs in a planar configuration on Au, thereby increasing the areal density of nitrile oscillators within the gap[42,43]. IR spectroscopy confirms absorption peak in the silent zone (2000-2500cm$^{-1}$) corresponding to the -C≡N stretch in BPTCN and TCNQ monolayers (Fig. 3a), while this feature is absent for BPT. Thus, under 4.55µm excitation, BPT acts as a vibrationally off-resonant control, whereas BPTCN and TCNQ provide resonant vibrational donors with different oscillator densities and geometries. Fig. 4c shows a histogram of the MIRVAL efficiency measured at 30 different MIM rings for each donor molecules assembled in the gaps. The non-resonant control spacer BPT exhibits a $\%\zeta_{\mathrm{MIR-VAL}}$ of 0.08±0.04%, due to weak photothermal perturbed SERS and the signal is least three- to five-fold lower than BPTCN. The highest efficiency is obtained with TCNQ (0.57±0.22%), albeit with larger ring-to-ring variability. The larger variance observed for TCNQ reflects an additional structural contribution and gap thickness variations[43,44]. The suppression of MIR-VAL in BPT confirms that broadband heating or non-resonant electronic nonlinearities cannot account for the observed upconversion; resonance with a specific vibrational donor is required.

## Discussion

This work introduces a vibrationally mediated donor–acceptor mechanism, where MIR photons first populate a molecular vibrational state and are then transferred via plasmon-enhanced near fields to an electronic emitter before ultrafast intramolecular relaxation dissipates the energy, which is fundamentally different from recent observations[23,45–52]. The observed vibrationally mediated upconversion efficiencies (<1%) are lower than the ~10% previously reported in plasmonic systems[23] without an intermediate vibrational donor. In those architectures[23], the MIR field couples directly and the conversion efficiency is governed primarily by electronic transition probabilities and cavity Purcell enhancement. In contrast, the present architecture inserts a vibrational intermediate whose population must compete with ultrafast IVR. The upconversion efficiency is therefore fundamentally constrained by the ratio $\frac{\Gamma_{\mathrm{DA}}}{\Gamma_{\mathrm{IVR}}}$, where $\Gamma_{\mathrm{DA}}$ is the plasmon-mediated vibrational donor-acceptor transfer rate and $\Gamma_{\mathrm{IVR}}$ is the intrinsic intramolecular relaxation rate (typically 0.1–10 $ps^{-1}$). The sub-percent efficiencies reported here are consistent with a regime in where $\Gamma_{\mathrm{DA}} \lesssim \Gamma_{\mathrm{IVR}}$; that is, the plasmon-enhanced transfer channel is competitive but not yet dominant over intramolecular redistribution. The role of collective excitation of molecular vibrations and optomechanical coupling in the nanogaps need further exploration[53–58]. The molecular comparison (BPT, BPTCN, TCNQ) demonstrates that increasing oscillator density and reducing gap volume both push the system toward the regime $\Gamma_{\mathrm{DA}} \sim \Gamma_{\mathrm{IVR}}$, providing a clear pathway for optimisation.

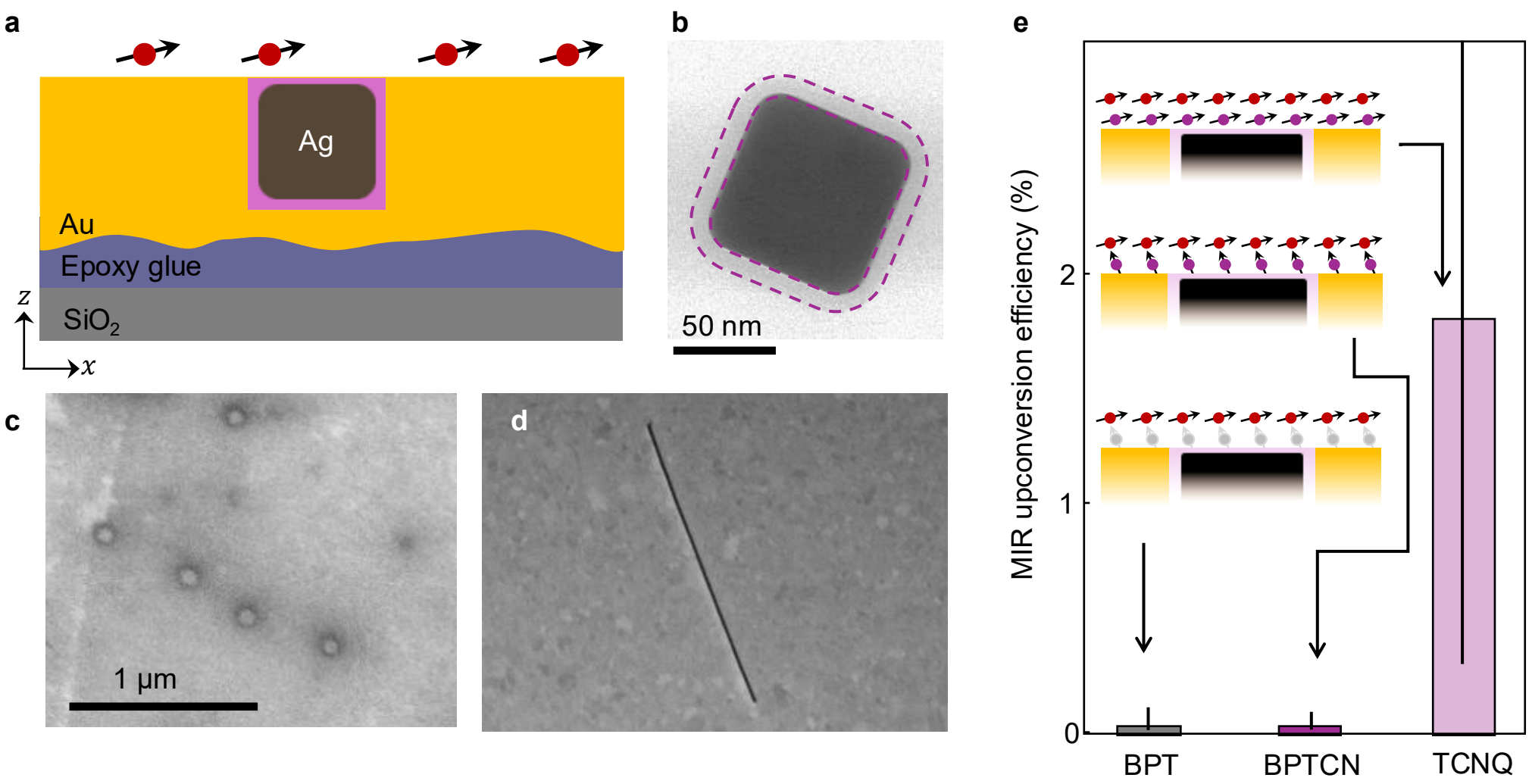

**Figure 4: Colloidal nanoparticle templates for MIM gaps for orientation-controlled donor–acceptor coupling.** (**a**) Cross-sectional schematic of a nanoparticle-templated MIM nanocube cavity with MB emitters assembled on the surface. (**b**) Scanning transmission electron microscopy image of an Ag nanocube surrounded by a polyvinylpyrrolidone (PVP) shell (dashed outline), which defines the nanoscale gap thickness upon assembly. (**c**) Scanning electron microscopy image of template-stripped MIM nanocube structures, forming discrete MIM nanocube cavity. (**d**) SEM image of an MIM nanowire cavity, providing an elongated gap geometry. (**e**) Histogram of MIR upconversion efficiency for MIM nanocube cavity functionalized with different donor-acceptor configurations (BPT, BPTCN, TCNQ). Error bars indicate the standard error. The red dot arrows indicate the orientation of MB electronic dipole and pink dot arrow indicates the vibrational dipole of -C≡N. The efficiency depends strongly on both the presence of the -C≡N vibrational donor and the relative orientation of donor and acceptor dipoles within the vertically confined plasmonic field.

The observed variation in upconversion efficiency across MIM rings originates primarily from nanoscale roughness along the vertical slot walls. Because $\Gamma_{\mathrm{DA}} \propto 1/V_{\mathrm{mode}}$, small fluctuations in gap thickness along the z-direction lead to large variations in local density of optical states and transfer rate. Achieving atomically flat vertical interfaces is therefore essential for reproducible confinement. This remains challenging in both top-down lithography and bottom-up template processes[32]. To mitigate this limitation, we provide next directions with colloidally synthesized nanostructures as gap templates (Fig.4). Silver nanocubes with atomically flat {100} vertical crystalline facets and a ~2nm polymer shell (Fig.4a,b) forms ideal templates for MIM gaps. Thermal evaporation of Au followed by template stripping produces MIM nanogaps conformal to the cube geometry, with the polymer shell defining the gap thickness (Supplementary Fig.S2). FDTD simulations of 75nm cubes show maximum field enhancement localized at cube edges, with peak enhancement factors ~27 in the MIR. Systematic variation of cube length tunes the resonance position and confinement strength (Supplementary Fig.S12c). These structures provide improved geometric control and a platform for further optimization of vibrationally mediated upconversion. Similarly, we also fabricated samples with templating silver nanowires to 1-D MIM gaps (Fig.4d), providing routes of further optimisation and exploration for strong vibrational donor molecules and efficient dye emitters.

Importantly, the MIM nanocube geometry allows us to decouple gap size from dipolar orientation effects. While the gap thickness remains fixed by the PVP spacer, the orientation of the vibrational donor dipole now changes depending on molecular assembly (Fig.4e inset). BPTCN forms a thiolated SAM on Au with its -C≡N bonds oriented predominantly perpendicular to the metal surface, aligning the vibrational dipole largely normal to the substrate. In contrast, TCNQ assembles by π-stacking parallel to the metal film, positioning its -C≡N dipoles primarily in-plane. These two configurations create opposite dipole-dipole coupling scenarios relative to the plasmonic field and vibrational transitions of MB. Strikingly, BPTCN-functionalized MIM cavities exhibit a near-complete suppression of MIR upconversion ($\%\zeta_{\mathrm{MIRVAL}}$=0.03 ± 0.01%), in sharp contrast to the MIM ring geometry where in-plane -C≡N bonds are optimally aligned with the plasmonic field and MB vibrations. Conversely, TCNQ produces enhanced $\%\zeta_{\mathrm{MIRVAL}}$>1% that exceeds those of the MIM rings, consistent with improved dipolar overlap in the MIM nanocube geometry. However, there is still large variability in $\%\zeta_{\mathrm{MIRVAL}}$ across the MIM gaps this likely arises from random molecular assembly, local roughness of the PVP spacer (Supplementary Fig.S12b), and possible chemical interactions between TCNQ and Au that may introduce additional charge-transfer contributions.

To rationalise the vibrational energy transfer process in plasmonic nanogaps, we develop a rate-equation model that captures the competition between IVR and plasmon-mediated donor-acceptor coupling. In plasmonic nanocavities, $\Gamma_{DA}$arises from a two-step process involving donor–plasmon and plasmon–acceptor coupling, and is therefore governed by both near-field enhancement and plasmon loss. In the weak-coupling regime, this leads to an effective rate $\Gamma_{DA} \propto | E_D/E_0 |^2 | E_A/E_0 |^2/(\kappa V_{\mathrm{eff}}^2)$, where $| E_{D,A}/E_0 |^2$ are the local field enhancements at the donor and acceptor positions, $V_{\mathrm{eff}}$ is the effective mode volume, and $\kappa$is the plasmon decay rate (Supplementary Section. 2). This motivates a plasmonic figure of merit for vibrational energy transfer,

$$\mathrm{FOM_{VET}} = \frac{| E_D/E_0 |^2 \quad | E_A/E_0 |^2}{\kappa\, V_{\mathrm{eff}}^2},$$

which captures the key requirement of simultaneously maximising near-field coupling at both molecules while minimising dissipative loss. In this framework, the upconverted signal scales as $I_{\mathrm{MIRVAL}} \propto \frac{k_{ve}\Gamma_{DA}}{\Gamma_D + k_{DA}}$, where $\Gamma_D$ ($\equiv \Gamma_{\mathrm{IVR}}$) is the total donor relaxation rate and $k_{ve}$ is the rate of conversion into an emissive electronic state. In electronic systems, efficient energy transfer is achieved because donor lifetimes (∼ns) are long compared to transfer rates ($\sim 10^9$-$10^{10}\mathrm{s}^{-1}$), yielding efficiencies approaching unity[59–62]. In contrast, vibrational excitations decay on sub-picosecond timescales ($\sim 10^{12}\mathrm{s}^{-1}$), requiring transfer rates that are orders of magnitude faster to compete. Thus, the scaling of this process is very steep with $\zeta \propto 1/V_{\mathrm{eff}}^2$ (Supplementary Section.2.3). This highlights the central requirement for vibrational energy transfer: the plasmon-assisted transfer rate must compete with or exceed IVR ($\Gamma_{DA} \gtrsim \Gamma_D$). The model further predicts strong sensitivity to spectral overlap, molecular orientation, and nanogap confinement, providing a quantitative framework to interpret the observed dependence of upconversion efficiency on molecular species and cavity geometry.

In conclusion, we demonstrate continuous-wave mid-infrared-to-visible upconversion mediated by vibrational donor-acceptor transfer within a plasmonic nanogap. By defining the MIM rings with a self-assembled molecular monolayer, we simultaneously achieve extreme electromagnetic confinement and introduce a resonant vibrational donor. The geometry is engineered to support strongly confined lateral gap fields that couple efficiently to in-plane molecular dipoles, maximising light-matter interaction and transfer rates. MIR excitation of the -C≡N stretch generates a non-equilibrium vibrational population that, inside the nanocavity, transfers energy to a nearby electronic acceptor before intramolecular vibrational redistribution dissipates it, resulting in enhanced anti-Stokes emission. Molecular comparisons confirm that the efficiency scales with resonant oscillator density and mode confinement, consistent with competition between plasmon-enhanced transfer ($\Gamma_{\mathrm{DA}}$) and vibrational relaxation ($\Gamma_{\mathrm{IVR}}$). Although the conversion efficiency remains sub-percent, the architecture operates entirely at room temperature and converts spectrally selective MIR absorption into visible emission, enabling detection without semiconductor dark-current limitations. These results demonstrate a vibrational analogue of donor-acceptor energy transfer in the MIR and provide a scalable route to study vibrational conformations, interaction dyanmcis, MIR sensing and cavity-enhanced molecular optomechanics.

## Materials and Methods

**Sample preparation.** Metal–molecule–metal ring resonators were fabricated using polystyrene microspheres of three different diameters: 500 nm (Sigma Aldrich, 59769), 7 µm (Sigma Aldrich, 78462), and 30 µm (Sigma Aldrich, 84135). Silicon wafers were sequentially cleaned with acetone, isopropyl alcohol, and deionised (DI) water, followed by oxygen plasma treatment (Diener Electronic plasma cleaner) for 5 min to render the surface hydrophilic. A diluted microsphere suspension (1 µL microspheres in 1 mL DI water) was drop-cast onto the silicon wafer and allowed to dry for 1 h, enabling self-assembly of the microspheres on the hydrophilic surface. To reduce aggregation of the microspheres, the sample was briefly exposed to oxygen plasma for 1 min. A 100 nm gold layer was subsequently deposited by thermal evaporation using Cresington Coating System 308R. The deposition took place at a base pressure of $10^{-6}$ mbar with a deposition rate of 0.1 nm/s, and source to substrate distance of approximately 15 cm. The substrate was kept at the room temperature during deposition and the film was deposited at a normal incidence. The microspheres were then completely removed by oxygen plasma etching at 300W power for 10 min, resulting in ring-shaped structures. The samples were immersed in molecular solutions for 12 h to form self-assembled monolayers (SAMs). Three different molecules were used for comparative analysis: BPT (Sigma Aldrich, 752207), BPTCN (Sigma Aldrich, 760080), and TCNQ (Sigma Aldrich, 157635). After SAM formation, a second gold layer of 200 nm thickness was deposited by thermal evaporation. Square glass substrates (1 cm × 1 cm) were coated with epoxy and attached to the wafer. The assembly was exposed to UV light for 30 min to cure the epoxy and left overnight to ensure complete curing. Finally, the glass substrates were mechanically stripped from the wafer, yielding an ultra-flat gold surface containing metal–molecule–metal ring structures. The sample was immersed in methylene blue (Sigma Aldrich, M6900) solution for 3 h before performing the Raman measurements. The schematic of the fabrication step is shown in Supplementary (Fig.S1).

Metal-insulator-metal (MIM) nanocube structures were fabricated using silver nanocubes (Nano Composix, KJW1966) with a side length of 75nm coated with a ~2nm polyvinylpyrrolidone (PVP) polymer layer. Silicon wafers were sequentially cleaned with acetone, isopropyl alcohol, and deionized (DI) water, followed by oxygen plasma treatment to render the surface hydrophilic. A diluted nanocube suspension (1µL nanocubes in 1mL DI water) was drop-cast onto the silicon wafer and allowed to dry for 1 h, enabling self-assembly of the nanocubes on the hydrophilic surface. A 100 nm thick gold layer was subsequently deposited by thermal evaporation. Square glass substrates (1 cm × 1 cm) were coated with epoxy and attached to the wafer surface. The assembly was exposed to UV light for 30 min to cure the epoxy and left overnight to ensure complete curing. The glass substrates were then mechanically stripped from the wafer, resulting in an ultra-flat gold surface containing embedded metal-insulator-metal nanocube structures. The samples were subsequently immersed in molecular solutions (BPT, BPTCN, or TCNQ) for 12h to form self-assembled monolayers (SAMs). After SAM formation, the samples were rinsed with DI water to remove excess molecules and then immersed in a methylene blue solution for 3h. The schematic of the fabrication step is shown in Supplementary (Fig.S2).

**Experimental setup.** Raman scattering measurements: All the Raman spectroscopy measurements were performed in a custom-built dual-channel microscope integrated with Renishaw spectrometer. For SERS, a spectrally filtered 785 nm diode laser with a power of 100 $\mu W/\mu m^2$ on the sample was used as a probe. It was filtered with two notch filters before being routed to a spectrograph and CCD. The 785nm light is focused onto the sample with the aid of a 100×, 0.9 NA short-working-distance microscope objective. For the MIR light source, a Thorlabs Turnkey Fabry- Pérot Quantum Cascade Laser was used. The pump (MIR light) was co-aligned with the probe (NIR light) using 15x reflective objective lens. The schematic of the fabrication step is shown in Supplementary Fig.S10.

Reflection measurements: FTIR measurements were performed using the Bruker Lumos FTIR microscope in reflection mode, equipped with an 8× objective lens and a liquid nitrogen-cooled mercury cadmium telluride (LN-MCT) detector, with 64 scans.

SEM images: SEM images were captured using an Apreo 2 Scanning Electron Microscope of Thermofisher Scientific.

TEM images: TEM images were captured using a 200kV JEOL2100 Scanning Transmission Electron Microscope.

**Acknowledgements**

We acknowledge funding from UKRI Future Leaders Fellowship, EPSRC (EP/Y008774/1) and the Royal Society (RGS/R1/231458). A.P acknowledges support from Commonwealth Split-site Scholarship.

**Author Information**

**Corresponding Author**

* Dr Rohit Chikkaraddy, r.chikkaraddy@bham.ac.uk

**Author Contributions**

R.C conceived and designed the experiments. A.P. performed the experiments with input from C.S and A.S; A.S performed FTIR measurements and simulation; E.A and W.T performed scanning transmission microscopy. R.C. and A.P. wrote the manuscript with input from all authors.

**Conflict of Interest**

The authors declare no competing financial interest.

**Supporting Information**. A Supporting Information document is also provided, with additional images and information. Source data can be found at https:// (to be provided on acceptance).

**References**:

1. Főrster, T. 10th Spiers Memorial Lecture. Transfer mechanisms of electronic excitation. *Discuss. Faraday Soc.* **27**, 7–17 (1959).
2. Dexter, D. L. A Theory of Sensitized Luminescence in Solids. *J. Chem. Phys.* **21**, 836–850 (1953).
3. Yang, F., Sambles, J. R. & Bradberry, G. W. Long-range coupled surface exciton polaritons. *Phys. Rev. Lett.* **64**, 559–562 (1990).

4. Andrews, D. L. & Demidov, A. A. *Resonance Energy Transfer*. (Wiley, 1999).
5. Scholes, G. D. Long-Range Resonance Energy Transfer in Molecular Systems. *Annual Review of Physical Chemistry* **54**, 57–87 (2003).
6. Engel, G. S. *et al.* Evidence for wavelike energy transfer through quantum coherence in photosynthetic systems. *Nature* **446**, 782–786 (2007).
7. Moerner, W. E. (William E. ). Nobel Lecture: Single-molecule spectroscopy, imaging, and photocontrol: Foundations for super-resolution microscopy. *Rev. Mod. Phys.* **87**, 1183–1212 (2015).
8. Su, R. *et al.* FRET Materials for Biosensing and Bioimaging. *Chem. Rev.* **125**, 9429–9551 (2025).
9. Roy, R., Hohng, S. & Ha, T. A practical guide to single-molecule FRET. *Nat Methods* **5**, 507–516 (2008).
10. Chen, T.-T., Du, M., Yang, Z., Yuen-Zhou, J. & Xiong, W. Cavity-enabled enhancement of ultrafast intramolecular vibrational redistribution over pseudorotation. *Science* **378**, 790–794 (2022).
11. Fernández-Terán, R. & Hamm, P. A closer look into the distance dependence of vibrational energy transfer on surfaces using 2D IR spectroscopy. *J. Chem. Phys.* **153**, 154706 (2020).
12. Gruebele, M. & Wolynes, P. G. Vibrational Energy Flow and Chemical Reactions. *Acc. Chem. Res.* **37**, 261–267 (2004).
13. Nesbitt, D. J. & Field, R. W. Vibrational Energy Flow in Highly Excited Molecules: Role of Intramolecular Vibrational Redistribution. *J. Phys. Chem.* **100**, 12735–12756 (1996).
14. He, H. *et al.* Mapping enzyme activity in living systems by real-time mid-infrared photothermal imaging of nitrile chameleons. *Nat Methods* **21**, 342–352 (2024).
15. Yin, J. *et al.* Video-rate mid-infrared photothermal imaging by single-pulse photothermal detection per pixel. *Science Advances* **9**, eadg8814 (2023).
16. Wang, H. *et al.* Bond-selective fluorescence imaging with single-molecule sensitivity. *Nat. Photon.* **17**, 846–855 (2023).
17. Huang, K., Fang, J., Yan, M., Wu, E. & Zeng, H. Wide-field mid-infrared single-photon upconversion imaging. *Nat Commun* **13**, 1077 (2022).
18. Barh, A., Rodrigo, P. J., Meng, L., Pedersen, C. & Tidemand-Lichtenberg, P. Parametric upconversion imaging and its applications. *Adv. Opt. Photon., AOP* **11**, 952–1019 (2019).
19. Wang, Y. *et al.* Mid-Infrared Single-Photon Edge Enhanced Imaging Based on Nonlinear Vortex Filtering. *Laser & Photonics Reviews* **15**, 2100189 (2021).
20. Gemmell, N. R. Loss-Compensated and Enhanced Midinfrared Interaction-Free Sensing with Undetected Photons. *Phys. Rev. Appl.* **19**, (2023).
21. Yao, J. *et al.* Giant single-step upconversion via sub–35-fs phonon dynamics in the nonlinear optical regime. *Science Advances* **11**, eadx1686 (2025).
22. Quan, J. *et al.* On-site enhancement and control of spin-forbidden dark excitons in a plasmonic heterostructure. *Nat. Photon.* **20**, 49–54 (2026).
23. Chikkaraddy, R., Arul, R., Jakob, L. A. & Baumberg, J. J. Single-molecule mid-infrared spectroscopy and detection through vibrationally assisted luminescence. *Nat. Photon.* 1–7 (2023) doi:10.1038/s41566-023-01263-4.
24. Baumberg, J. J., Aizpurua, J., Mikkelsen, M. H. & Smith, D. R. Extreme nanophotonics from ultrathin metallic gaps. *Nature Materials* **18**, 668–678 (2019).
25. Li, Y. *et al.* Boosting Light−Matter Interactions in Plasmonic Nanogaps. *Advanced Materials* **36**, 2405186 (2024).

26. Xomalis, A. *et al.* Detecting mid-infrared light by molecular frequency upconversion in dual-wavelength nanoantennas. *Science* **374**, 1268–1271 (2021).
27. Lee, J. *et al.* Extraordinary optical transmission and second harmonic generation in sub–10-nm plasmonic coaxial aperture. *Nanophotonics* **9**, 3295–3302 (2020).
28. Yoo, D. *et al.* High-Contrast Infrared Absorption Spectroscopy via Mass-Produced Coaxial Zero-Mode Resonators with Sub-10 nm Gaps. *Nano Lett.* **18**, 1930–1936 (2018).
29. Jiang, K. *et al.* Large-Scale Fabrication of 5 nm Plasmonic Hybrid Nanoslit Arrays. *Nano Lett.* **25**, 8636–8643 (2025).
30. Beesley, D. J. *et al.* Sub-15-nm patterning of asymmetric metal electrodes and devices by adhesion lithography. *Nat Commun* **5**, 3933 (2014).
31. Namgung, S., Koester, S. J. & Oh, S.-H. Ultraflat Sub-10 Nanometer Gap Electrodes for Two-Dimensional Optoelectronic Devices. *ACS Nano* https://doi.org/10.1021/acsnano.0c10759 (2021) doi:10.1021/acsnano.0c10759.
32. Yoo, D. *et al.* Modeling and observation of mid-infrared nonlocality in effective epsilon-near-zero ultranarrow coaxial apertures. *Nat Commun* **10**, 4476 (2019).
33. Shi, J. *et al.* A room-temperature polarization-sensitive CMOS terahertz camera based on quantum-dot-enhanced terahertz-to-visible photon upconversion. *Nat. Nanotechnol.* **17**, 1288–1293 (2022).
34. Baiz, C. R. *et al.* Vibrational Spectroscopic Map, Vibrational Spectroscopy, and Intermolecular Interaction. *Chem. Rev.* **120**, 7152–7218 (2020).
35. Rubtsova, N. I. & Rubtsov, I. V. Vibrational Energy Transport in Molecules Studied by Relaxation-Assisted Two-Dimensional Infrared Spectroscopy. *Annual Review of Physical Chemistry* **66**, 717–738 (2015).
36. Chikkaraddy, R. *et al.* Mapping Nanoscale Hotspots with Single-Molecule Emitters Assembled into Plasmonic Nanocavities Using DNA Origami. *Nano Lett.* **18**, 405–411 (2018).
37. Kongsuwan, N. *et al.* Suppressed Quenching and Strong-Coupling of Purcell-Enhanced Single-Molecule Emission in Plasmonic Nanocavities. *ACS Photonics* **5**, 186–191 (2018).
38. Horton, M. J. *et al.* Nanoscopy through a plasmonic nanolens. *PNAS* **117**, 2275–2281 (2020).
39. Rocchetti, S. *et al.* Amplified Plasmonic Forces from DNA Origami-Scaffolded Single Dyes in Nanogaps. *Nano Lett.* **23**, 5959–5966 (2023).
40. Griffiths, J. *et al.* Resolving sub-angstrom ambient motion through reconstruction from vibrational spectra. *Nat Commun* **12**, 6759 (2021).
41. Carnegie, C. *et al.* Room-Temperature Optical Picocavities below 1 nm3 Accessing Single-Atom Geometries. *J. Phys. Chem. Lett.* **9**, 7146–7151 (2018).
42. Kang, J. F. *et al.* Self-Assembled Rigid Monolayers of 4′-Substituted-4-mercaptobiphenyls on Gold and Silver Surfaces. *Langmuir* **17**, 95–106 (2001).
43. Otero, R., Miranda, R. & Gallego, J. M. A Comparative Computational Study of the Adsorption of TCNQ and F4-TCNQ on the Coinage Metal Surfaces. *ACS Omega* **4**, 16906–16915 (2019).
44. Patterson, T., Pankow, J. & Armstrong, N. Tetracyanoquinodimethane thin films on copper, gold, platinum, and tin (IV) sulfide: characterization by x-ray photoelectron spectroscopy. *Langmuir* **12**, 3160–6 (1991).
45. Whaley-Mayda, L., Guha, A., Penwell, S. B. & Tokmakoff, A. Fluorescence-Encoded Infrared Vibrational Spectroscopy with Single-Molecule Sensitivity. *J. Am. Chem. Soc.* **143**, 3060–3064 (2021).

46. Chikkaraddy, R., Xomalis, A., Jakob, L. A. & Baumberg, J. J. Mid-infrared-perturbed molecular vibrational signatures in plasmonic nanocavities. *Light Sci Appl* **11**, 19 (2022).
47. Liang, L., Wang, C., Chen, J., Wang, Q. J. & Liu, X. Incoherent broadband mid-infrared detection with lanthanide nanotransducers. *Nat. Photon.* **16**, 712–717 (2022).
48. Meng, Z.-D. *et al.* Colocalized Raman and IR Spectroscopies via Vibrational-Encoded Fluorescence for Comprehensive Vibrational Analysis. *J. Am. Chem. Soc.* **147**, 16309–16318 (2025).
49. Zhang, Y. *et al.* Fluorescence-Detected Mid-Infrared Photothermal Microscopy. *J. Am. Chem. Soc.* **143**, 11490–11499 (2021).
50. Wang, C. W. *et al.* Mid-infrared detection through ligand-driven local heating in lanthanide-doped nanoparticles. *Nat Commun* https://doi.org/10.1038/s41467-026-70900-7 (2026) doi:10.1038/s41467-026-70900-7.
51. Wang, H. *et al.* Room-Temperature Single-Molecule Infrared Imaging and Spectroscopy through Bond-Selective Fluorescence. *Angewandte Chemie International Edition* **63**, e202413647 (2024).
52. Qian, N., Xiong, H., Wei, L., Shi, L. & Min, W. Merging Vibrational Spectroscopy with Fluorescence Microscopy: Combining the Best of Two Worlds. *Annual Review of Physical Chemistry* **76**, 279–301 (2025).
53. Puro, R. L., Gray, T. P., Kapfunde, T. A., Richter-Addo, G. B. & Raschke, M. B. Vibrational Coupling Infrared Nanocrystallography. *Nano Lett.* **24**, 1909–1915 (2024).
54. Dows, D. A. Intermolecular Coupling of Vibrations in Molecular Crystals. *J. Chem. Phys.* **32**, 1342–1347 (1960).
55. Wilcken, R. *et al.* Antenna-coupled infrared nanospectroscopy of intramolecular vibrational interaction. *Proceedings of the National Academy of Sciences* **120**, e2220852120 (2023).
56. Jakob, L. A. *et al.* Optomechanical Pumping of Collective Molecular Vibrations in Plasmonic Nanocavities. *ACS Nano* **19**, 10977–10988 (2025).
57. Loirette-Pelous, A., Boto, R. A., Aizpurua, J. & Esteban, R. Addressing intramolecular vibrational redistribution in a single molecule through pump and probe surface-enhanced vibrational spectroscopy. Preprint at https://doi.org/10.48550/arXiv.2601.02117 (2026).
58. Shalabney, A. *et al.* Coherent coupling of molecular resonators with a microcavity mode. *Nat Commun* **6**, (2015).
59. Hildner, R., Brinks, D., Nieder, J. B., Cogdell, R. J. & Van Hulst, N. F. Quantum coherent energy transfer over varying pathways in single light-harvesting complexes. *Science* **340**, 1448–1451 (2013).
60. Ghenuche, P. *et al.* Matching Nanoantenna Field Confinement to FRET Distances Enhances Förster Energy Transfer Rates. *Nano Lett.* **15**, 6193–6201 (2015).
61. Hamza, A. O., Viscomi, F. N., Bouillard, J.-S. G. & Adawi, A. M. Förster Resonance Energy Transfer and the Local Optical Density of States in Plasmonic Nanogaps. *J. Phys. Chem. Lett.* **12**, 1507–1513 (2021).
62. Xie, X. *et al.* Plasmonic Nanocavity-Assisted Long-Range Dipole–Dipole Interactions for Rare-Earth Ions. *Nano Lett.* **26**, 2877–2885 (2026).